\documentclass[preprint,12pt]{elsarticle}
\usepackage{amssymb}

\journal{Advanced Powder Technology}

\begin{document}
\begin{frontmatter}

\title{Simulation of grains in a vibrated U-tube without interstitial fluid}

\author[usb]{J.R. Darias}
\author[ivic]{I.  S\'{a}nchez}
\author[usb]{G. Guti\'{e}rrez}
\author[ani]{R. Paredes}

\address[usb]{Departamento de F\'{i}sica, Universidad Sim\'{o}n Bol\'{i}var, Apartado Postal 89000, Caracas 1080-A, Venezuela.}
\address[ivic]{Laboratorio de F\'{i}sica Estad\'{i}stica de Medios Desordenados, Centro de F\'{i}sica, Instituto Venezolano de Investigaciones Cient\'{i}ficas, Apartado Postal 21827, Caracas 1020-A, Venezuela.}
\address[ani]{Academia Nacional de Investigaci\'{o}n y Desarrollo A.C., C\'{o}digo Postal 62040, Palmira N°. 13, Cuernacava, Morelos, M\'{e}xico.}

\begin{abstract}
We present a computational study using Molecular Dynamics of the development of an accumulation of grains in one side of a two dimensional U-tube under vertical vibrations. Studying the evolution of the height difference between the centers of mass of the branches of the tube, we found that it reaches a saturation value after vibrating for some time. We obtain that this saturation value is the same if the simulation start with the arms leveled or with a large initial height difference. We explore the effect of the width of the tube, the density of the grains and the coefficient of restitution between the grains and the wall on the value of the saturation. We obtain a value of the width of the tube for which the saturation value reaches a maximum, and show that the transport of grains between arms is favored for low grain densities and high grain-wall restitution coefficient.    
\end{abstract}


\end{frontmatter}

\section{Introduction}

Vertical vibrations can induce a collective motion of grains inside a U-shaped tube or a partitioned rectangular container. The main consequence of this collective motion is the accumulation of the granular material in one side of the tube, that can depend on several parameters like the shape of the container, the properties of the granular material, the frequency and amplitude of oscillation and the presence of any interstitial fluid. When the effect of an interstitial fluid between grains is relevant, the height difference between the limbs of a vibrated U-tube tends to grow exponentially. This behaviour has been studied experimentally for small grains ($\approx 200$ $\mu$m diameter) in air \cite{Darias2011,Ivan2009,Perez2011} and theoretically, computationally and experimentally for coarser grains ($\approx 800$ $\mu$m diameter) in water \cite{Clement2009}.

Akiyama et al. \cite{Akiyama2001} instead of using an U-tube, studied experimentally a rectangular container partitioned by a vertical wall on its center, with a small gap below for communication between both sides of the container. They observed that vibration on an initially leveled partitioned container, induced the accumulation of grains on one side of the container, in a way that the level difference between the partitions grew up to a certain height. They showed that the maximum height difference achieved was larger when the experiment was performed at atmospheric pressure, but shrank to a value of $\approx$ 20 grain diameters, when the air pressure was reduced below $50$ Torr. They also presented a simple snapshot of a two-dimensional molecular dynamics simulation of soft spheres, showing a maximum height difference in accordance with the experimental result at reduced air pressure, however no further information is given regarding the simulations. 

Ohtsuki et al. \cite{Ohtsuki1998}, studied a partitioned container like Akiyama et al. but with unequal partition widths. They also showed simulations to complement their experimental results, using two-dimensional polydisperse hard discs. In their simulations, the accumulation of grains always occurs towards the widest partition, while it can occur towards any side in symmetrically partitioned containers or U-tubes. They focused their computational study on how the kinetic friction between the walls and grains affects the transport of grains from one partition to another.

Until now, previous research has identified two key factors affecting the development of a height difference in partitioned containers or U-tubes: Energy loss due the grain-wall interaction and the presence of an interstitial fluid between grains. With the help of Molecular Dynamics simulation \cite{Allen1989,Rapaport2004} we decided to study the problem in the absence of an interstitial fluid, and in a simple two dimensional case, to characterize the problem in a controlled basic configuration. We studied the evolution of the height difference of the center of mass of the U-tube arms $\Delta y$, and also the dependence of the saturation value $\Delta y_{sat}$, as a function of the tube width, the mass of the grains and the coefficient of restitution between the walls and the grains.

\section{Computational model}

\begin{figure}[t]
\begin{center}
\includegraphics[width=4.0cm, height=8.0cm ]{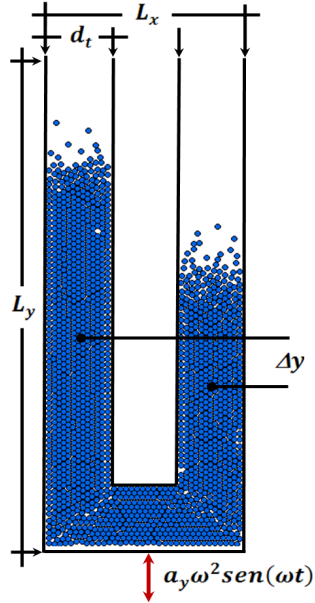}
\caption{Model of the granular material inside a U-tube under vibration. }\label{graph-0} 
\end{center}
\end{figure}

Our model consist of $N$ monosized discs with mass $m$ and diameter $d$, partially filling a thin U shaped tube with vertical branches of width $d_t$, height $L_y$ and external width $L_x$ (see Fig. \ref{graph-0}). The tube is vertically vibrated applying an acceleration of the form $a_yw^2sen(\omega t)$, where $a_y$ is the amplitude of vibration, $w$ is the angular frequency and $t$ is the time.

In our system, two grains $i$ and $j$ with positions $\{\vec{r}_i$, $\vec{r}_j\}$ and translational velocities $\{\vec{v}_i$, $\vec{v}_j\}$, will suffer an inelastic collision only when the separation between their centers of mass is lower than their effective diameter $r_{ij}=|\vec{r}_i-\vec{r}_j| < (d_i+d_j)/2$. During the collision, the $i$th grain will feel a restorative and dissipative contact force due to the $j$th grain, with normal component (through the line connecting their centers of mass, $\hat{n}$ direction) and tangential (through a line perpendicular to the line connecting their centers of mass, passing through the contact point, $\hat{s}$ direction) of the form:

\begin{center}
\begin{equation}\label{eqn-1}
\vec{F}_{ij}=\{k_n(d_{ij}-r_{ij})-\gamma_n m_rv^n_{ij}\}\hat{n}-\frac{\vec{v}_{ij}\cdot\hat{s}}{|\vec{v}^s_{ij}\cdot\hat{s}|}\,\{min(\gamma_s m_r|\vec{v}^s_{ij}|,\,\mu|\vec{F}_{ij}\cdot \hat{n}|)\}\hat{s},	
\end{equation}
\end{center}

\noindent where $\vec{v}_{ij}=\vec{v}_i-\vec{v}_j$, $\hat{n}= \vec{r}_{ij}/r_{ij}$, $v^n_{ij}=\vec{v}_{ij}\cdot\hat{n}$, $\hat{s}= \vec{v}^s_{ij}/|\vec{v}^s_{ij}|$ and $\vec{v}^s_{ij} = \vec{v}_{ij}-v^n_{ij}\hat{n}$.  The first term on the right represents the elastic grain-grain interaction acting during contact, the second term describes viscous forces for the normal deformation proportional to the collision velocities, and the last term accounts for the tangential friction between grains, where we take the minimum between viscous and Coulomb's friction. Thompson and Grest \cite{Thompson1991} found that rotations were irrelevant in two-dimensional modeling of a monosized granular material in the presence of gravity, therefore as first approach we do not include rotations in our model.

\begin{table}[th]
\label{table:parameters} 
\caption{Parameters used in simulations in units of $m$, $g$ and $d$.}
\begin{tabular}{l}
\hline
\hline
Number of grains \hspace{5.3cm} $610 - 2210$\\
Grain diameter \hspace{6.45cm} $1$ \\
Grain mass \hspace{6.3cm} $0.25 - 2.5$\\
Grain density \hspace{5.43cm}  $0.32\,\rho_0 - 3.16\,\rho_0$\\
Elastic constant $k_n$ \hspace{4.37cm}  $7989\,k_0 - 199992\,k_0$\\
Normal dissipation coefficient $\gamma_n$ \hspace{2.01cm} $7.98\,\gamma_0 - 1974\,\gamma_0$\\
Tangential dissipation coefficient $\gamma_s$ \hspace{1.50cm} $7.98\,\gamma_0 - 1974\,\gamma_0$\\
Friction coefficient $\mu$ \hspace{5.4cm} $0.5$\\
Grain-wall restitution coefficient $e_w$ \hspace{1.95cm} $0.05 - 1$\\
Tube width $d_t$ \hspace{5.75cm} $5.5\,d - 16.67\,d$\\
Tube height $L_y$ \hspace{6.0cm} $120.5\,d$\\ 
Base width $L_x$ \hspace{5.5cm} $23.1\,d - 45.44\,d$\\
Vibration amplitude $a_y$ \hspace{4.7cm} $6.83\,d$ \\
Vibration frequency $w$ \hspace{4.75cm} $0.49\,w_0$\\
Time step $\Delta t$ \hspace{6.4cm} $10^{-4}\,t_0$\\
\hline
\hline
\end{tabular}
\end{table}

The parameters $k_n$, $\gamma_{n,s}$ and $\mu$, are the elastic constant, the coefficients of viscous friction and the kinetic coefficient of solid friction and $m_{r} = m_im_j/(m_i+m_j)$ is the reduced mass. The corresponding contact force on the $j$th grain is given by the third Newton's Law $\vec{F}_{ji}=-\vec{F}_{ij}$. Under the gravitational field $\vec{g}$, the translational accelerations of the grains are determined by the second Newton's law, in terms of the total force acting on grain $i$: $\vec{F}^{tot}_i = m_i\vec{g}+\sum_j\vec{F}_{ij}$. To save computational time during the calculation of the interaction force between grains, we used the cell algorithm \cite{Rapaport2004}. Part of the energy lost during a collision is quantified through the restitution coefficient $\epsilon_n$, for the normal deformation, given by $\epsilon_n=e^{-\gamma_n t_{col}/2}$, where $t_{col}=\pi/(k_n/m-\gamma_n^2/4)^{1/2}$, is the collision time, and the elastic constant $k_n$, is chosen such that two grains do not overlap more than $1\%$ under the action of gravity. Taking this in consideration, in order to guarantee numerical stability our time step was taken as $\Delta t \sim t_{col}/100$.

The U-tube has rough walls, such that the normal and parallel components of the pre-collision $\vec{v}_i$ and post-collision $\vec{v}\,'_i$ grain velocities are given by $v\,'_{i\perp} = -e_w |\vec{v}_i|sen(\theta_R)$ and $v\,'_{i\parallel} = e_w|\vec{v}_i|\cos(\theta_R)$. Here $e_w$, is a restitution coefficient characterizing the energy loss during collision with the walls, $|\vec{v}_i|$ is the magnitude of the pre-collision velocity and $\theta_R \in [0, \pi]$ is a random angle \cite{Rapaport2004,Darias2002,Darias2003}.

Equations of motion are integrated using a second order Velocity Verlet scheme \cite{Verlet1967,Verlet1968} and the results are given in dimensionless quantities using the normalization parameters $m$, $g$ y $d$. Considering this, the unit for mass is $m_0=1$, acceleration $g_0=1$, diameter $d_0=1$, time $t_0=\sqrt{d/g}$, angular frequency $w_0=\sqrt{g/d}$, velocity $v_0=\sqrt{gd}$, force $F_0=mg$, elastic constant $k_0=mg/d$, coefficient of viscous friction $\gamma_0=m\sqrt{g/d}$ and density $\rho_0=m/d^2$. Table \ref{table:parameters}, lists this quantities and the values used in our simulations.  

\section{Results and discussion}

\begin{figure}[t]
\begin{center}
\includegraphics[width=11.0cm, height=5.0cm]{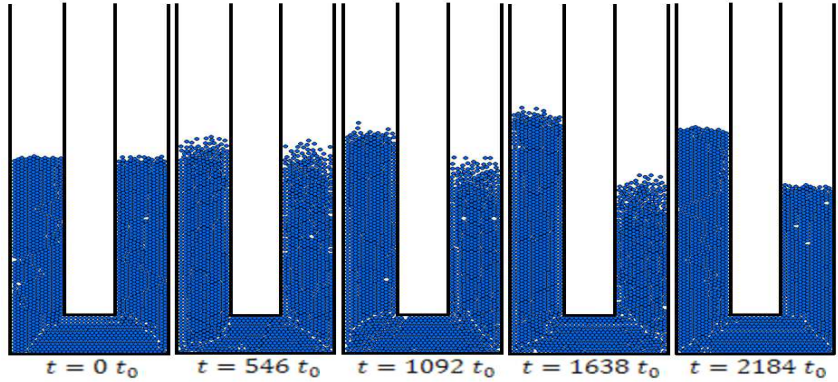}
\caption{Temporal evolution of $N=1700$ grains in a U-tube, filled up to $L_y/2$ with $L_x=35.7\,d$ and $d_t = 11\,d$. Starting with the arms leveled at time zero, the height difference $\Delta y$ between the centers of mass of each arm grows to a saturating value. The grain accumulation can happen in any arms of the tube, as it happens in the experiments at reduced air pressure reported by Akiyama et al \cite{Akiyama2001}.} \label{graph-1} 
\end{center}
\end{figure}

In all our simulations, the amplitude and frequency of the vertical acceleration are $a_y=6.83\,d $ and $\omega = 0.49\, \omega_0$ respectively. As we mentioned in the introduction, the height difference $\Delta y$ reaches a saturation value $\Delta y_{sat}$ after vertical vibration is applied. In Fig. \ref{graph-1} we shown the evolution the height difference $\Delta y$, where an accumulation of grains towards one side of the tube happens without any interstitial fluid, in a simple two dimensional setup and in the absence of an horizontal component in the oscillation.  

\begin{figure}[th]
\begin{center}
\includegraphics[width=10.0cm, height=7.0cm]{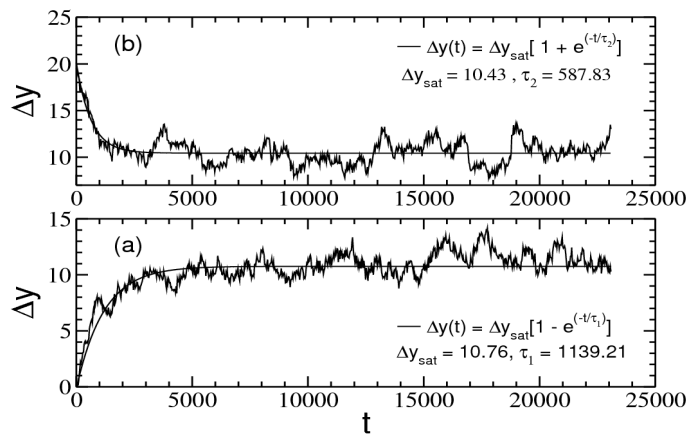}
\caption{ Evolution of $\Delta y$ for two different initial conditions. (a) Starting with the arms leveled. (b) Starting with an initial height difference of  $\Delta y_0= 20,86\, d$. Segmented lines are saturating exponential fits discussed in the text.}\label{graph-2}
\end{center}
\end{figure}

As we mentioned previously, the low accumulation of grains at reduced air pressure was initially observed in the experiments of Akiyama et al. \cite{Akiyama2001}. However, there is the reasonable doubt that the small accumulation observed by Akiyama et al. could be due to a small horizontal component present in the oscillation (since they do not discuss the issue of the verticality of the oscillation in their work). The presence of a small horizontal component can trigger the motion of grains to one side of a tube, as shown in reference \cite{Ivan2009}. The simulations of Ohtsuki et al. \cite{Ohtsuki1998}, in principle, solved this issue, because the oscillation is purely on the vertical direction, but their system is not symmetric since their partitioned container has one branch wider than the other. In our case, we observed the accumulation in a perfectly symmetric tube vibrated in a pure vertical way. This points out to the fact that the ingredient behind the accumulation is present in our simulations. It is also noticeable that the accumulation is observed in both the simulation of Ohtsuki et. al and ours, even with the different energy input mechanisms (Ohtsuki et al. shake the tube with a triangular waveform, while we shake used a sinusoid waveform) and different wall grain interaction (Ohtsuki et. al. simulated smooth walls, whereas we simulated rough walls).

The saturation value of $\Delta y$, is quite robust with respect to different initial conditions, if we begin the simulation with the arms leveled ($\Delta y_0 =0$), after a time $t=23400\, t_0$ $\Delta y$ grows and reaches a saturation value of $\Delta y_{sat} \approx 11\, d$ (see Fig. \ref{graph-2}a). If we start the simulations with the arms unbalanced ($\Delta y_0 =20.86\, d$), we found that $\Delta y$ decreases down to a saturating value of $\Delta y_{sat} \approx 11\, d$ (see Fig. \ref{graph-2}b) that is the same saturating value for the growing case.

\begin{figure}[!th]
\begin{center}
\includegraphics[width=10.0cm, height=7.0cm]{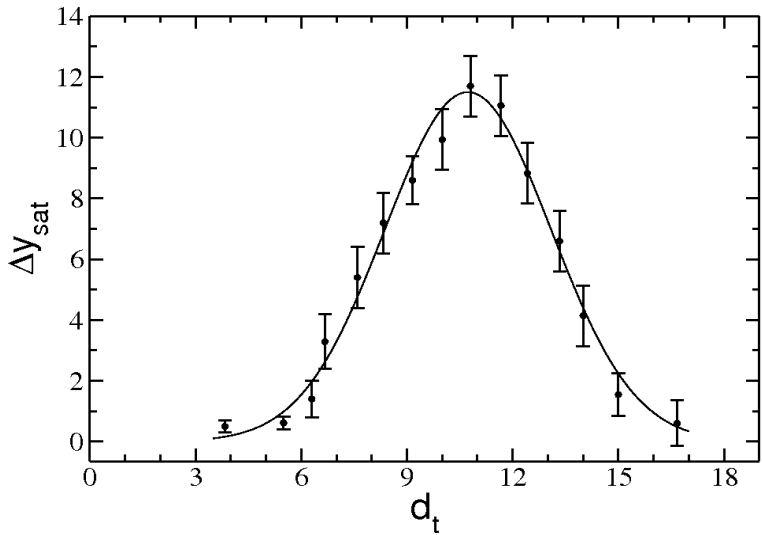}
\caption{Saturation height difference $\Delta y_{sat}$ as a function of the tube width $d_t$, showing a peak for intermediate widths. Each point is the average of ten runs started with the arms leveled. The parameters characterizing grain-grain and grain-wall interactions were fixed to $\rho = 1.27\, \rho_0$, $k_n = 20000\, k_0$, $\gamma_n = 20\, \gamma_0$, $\gamma_s = 20\, \gamma_0$, $e_w=0.9$.} \label{graph-3}
\end{center}
\end{figure}

Our results for the evolution of $\Delta y$ are well fitted by saturating exponential functions for both cases depicted in Fig. \ref{graph-2}. The growing case (a) is fitted by $\Delta y(t) = \Delta y_{sat}[1-e^{-t/\tau_1}]$ with $\Delta y_{sat}$ representing the saturation value reached and $\tau_1$ is the characteristic growth time. The decreasing case (b) is well fitted by $\Delta y(t) = \Delta y_{sat}[1+ e^{-t/\tau_2}]$. The fact that $\tau_1 > \tau_2$ can be understood since the transport of grains in the decreasing case is favored by gravity. The quantities $d/\tau_1\sim 10^{-4}$ and $d/\tau_2\sim 10^{-3}$, gives us an order of magnitude of the local effective velocity of a single grain moving in the tube, while $\Delta y_{sat}/\tau_1\sim 10^{-3}$ and $\Delta y_{sat}/\tau_2\sim 10^{-2}$ gives us an order of magnitude of the velocity of the bulk motion of the granular material through the tube. Both velocities are much lower than the velocity $v_0=\sqrt{gd}=1$ of a single grain in free fall after covering a distance equal to its own diameter. 

The saturation height $\Delta y_{sat}$ is the main focus of our subsequent results, since it is a effective measurement of the amount of accumulated grains in one side of the tube. The fact that the accumulation takes place up to a maximum height difference $\Delta y_{sat}$, that is the same whether the simulation is started with the arms of the tube leveled or with an initial height difference larger than $\Delta y_{sat}$, implies that the growing process of $\Delta y$ stops at the same point as the decreasing process, and further sustains the importance of the final height difference as a relevant parameter to monitor.

\begin{figure}[!th]
\begin{center}
\includegraphics[width=10.0cm, height=7.0cm]{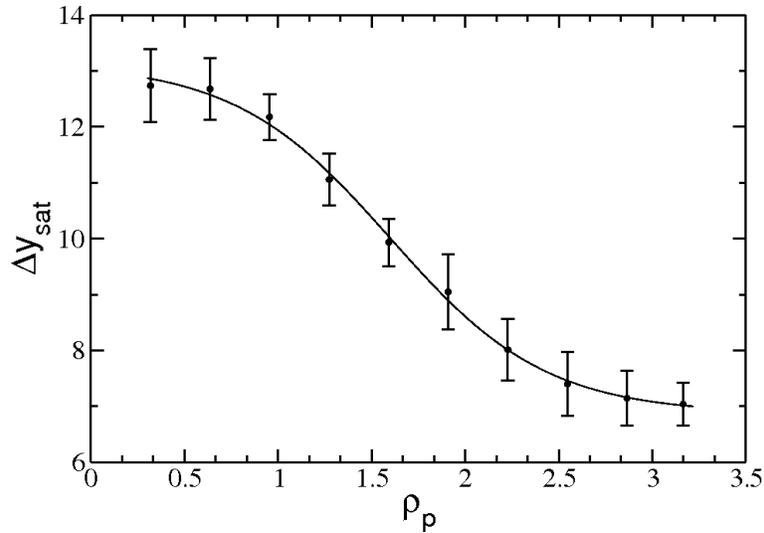}
\caption{Saturation height difference $\Delta y_{sat}$ as a function of the grain density, for a fixed tube width of $d_t=11\, d$, $k_n = 20000\, k_0$, $\gamma_n = 20\, \gamma_0$, $\gamma_s = 20\, \gamma_0$, $e_w=0.9$.  Each point is the average of ten runs started with the arms leveled.} \label{graph-4}
\end{center}
\end{figure}

Figure \ref{graph-3} shows the influence of the width of the tube $d_t$ on $\Delta y_{sat}$. Each point is obtained from several simulations starting with the arms leveled. For each tube width, the number of grains $N$ was adjusted so as to have a fixed initial height of grains in each branch of $L_y/2$. We can appreciate that the lower and larger tube widths ($d_t = 5.5\, d$ and  $d_t = 16.7\, d$) show practically no unbalance. This result suggest that there is an optimal tube width around $d_t=11\, d$ that maximize $\Delta y_{sat}$. This result can be understood since for very narrow tubes, the flow of grains may be severely affected by the probability of an obstruction occurring inside the tube, thus no accumulation develops for very narrow tubes. The fact that there is no accumulation in the other limit (large widths), is understood since in that case, the walls through a complex mechanism linked to the vibration and the friction force, cannot maintain the weight of the columns, therefore is not produced the accumulation of grains.

\begin{figure}[!th]
\begin{center}
\includegraphics[width=10.0cm, height=7.0cm]{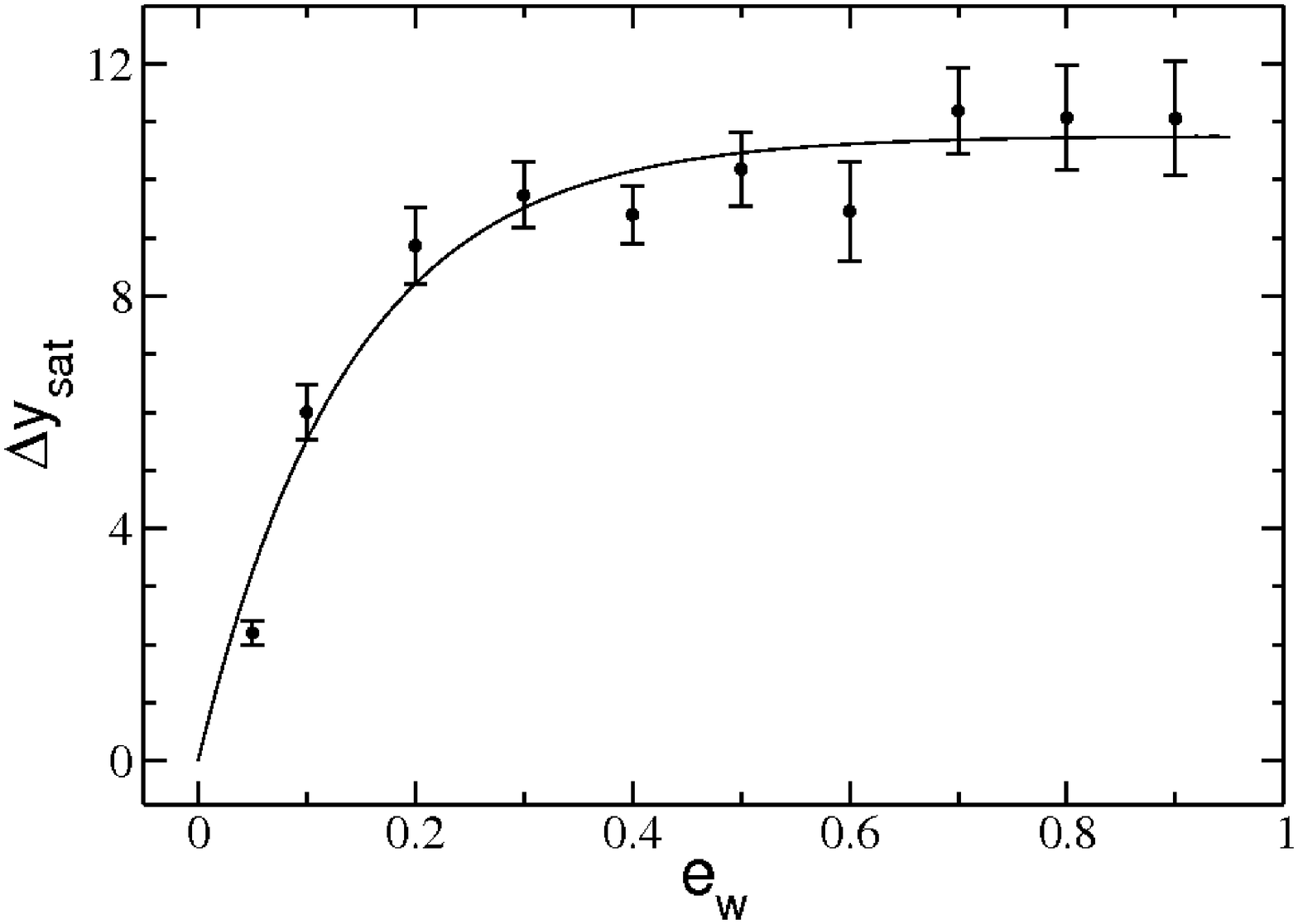}
\caption{$\Delta y_{sat}$ as a function of the grain-wall restitution coefficient $e_w$. These observation indicates that under conditions of large energy dissipation, the accumulation is frustrated. This results agrees with previous results reported in \cite{Ohtsuki1998,Landry2004}.} \label{graph-5}
\end{center}
\end{figure}

To probe the effect of the grain density on the accumulation, we plot in Fig. \ref{graph-4} the value of $\Delta y_{sat}$ for a fixed tube width varying the grain density to change the weight of the columns. The simulations were done using a tube width of $d_t=11\, d$, value around which the accumulative effect is more prominent, according to the results shown in figure \ref{graph-3}. The value of $\Delta y_{sat}$ decreases with increasing grain density. This result points out the fact that the accumulation process benefits from a low density granular material. At first glance, one is prompted to believe that the tube can hold the growing column only if its weight remains below a certain value. This behaviour is similar to an equilibrium condition, were the frictional force between wall and grains is not enough to hold up the granular column.  

To probe the effect of the wall on the growth of the columns, in Fig. \ref{graph-5} we show the variation of $\Delta y_{sat}$ with the restitution coefficient $e_w$. $\Delta y_{sat}$ decreases rapidly when $e_w$ goes to zero. This effect can be understood as follows, the grains that collide with the walls lose kinetic energy abruptly due the low value of $e_w$, these grains work like a ``strongly dissipative granular trap'' for the other grains, causing that lose energy quickly and stopping the accumulation process. For values of $e_w > 0.6$, $\Delta y_{sat}$ reaches an almost constant value. In this case, the effect of walls on the grains is poor and the energy lost is due practically to grain-grain interaction, which is not changed during the simulations.    

\section{Conclusion}

We studied the development of an accumulation of grains in one of the branches of a vertically vibrated two dimensional U-tube, in the absence of interstitial fluid and horizontal component. By studying the evolution of the height difference between the centers of mass of the arms of the tube $\Delta y$, we found that $\Delta y$ reaches a saturation value $\Delta y_{sat}$, that does not depend on the initial height difference. We probed the influence of the tube width, the grain density and the grain-wall coefficient of restitution on the development of the accumulation, by observing their effect on $\Delta y_{sat}$. We found that there is an optimal tube width maximizing the development of the accumulation, and that it does not develop for too narrow or too wide tubes. The unbalance is promoted for low grain densities and frustrated at larger grain densities. If the coefficient of restitution is small, the accumulation is also frustrated, but if the restitution coefficient is increased, there is a point over which $\Delta y_{sat}$ becomes independent of it.  
 
\bibliographystyle{elsarticle-num}
\bibliography{bibliography}

\end{document}